\documentclass[aps,amsmath,amssymb,pra,twocolumn,showpacs]{revtex4}
\usepackage{graphicx}

\begin{document}
\title{Casimir-Lifshitz Interaction between Dielectrics of
Arbitrary Geometry: A Dielectric Contrast Perturbation Theory}

\author{Ramin Golestanian}
\email{r.golestanian@sheffield.ac.uk} \affiliation{Department of
Physics and Astronomy, University of Sheffield, Sheffield S3 7RH,
UK}

\date{\today}

\begin{abstract}
The general theory of electromagnetic--fluctuation--induced
interactions in dielectric bodies as formulated by Dzyaloshinskii,
Lifshitz, and Pitaevskii is rewritten as a perturbation theory in
terms of the spatial contrast in (imaginary) frequency dependent
dielectric function. The formulation can be used to calculate the
Casimir-Lifshitz forces for dielectric objects of arbitrary
geometry, as a perturbative expansion in the dielectric contrast,
and could thus complement the existing theories that use
perturbation in geometrical features. We find that expansion in
dielectric contrast recasts the resulting Lifshitz energy into a sum
of the different many-body contributions. The limit of validity and
convergence properties of the perturbation theory is discussed using
the example of parallel semi-infinite objects for which the exact
result is known.
\end{abstract}


\maketitle

\section{Introduction}  \label{sec:intro}

Macroscopic material boundaries that interact with fluctuating
electromagnetic fields experience an induced interaction amongst
themselves. This was first demonstrated for the case of two
perfectly conducting parallel plates by Casimir \cite{Casimir48},
and was subsequently generalized to take into account frequency
dependent dielectric properties of the objects by Lifshitz
\cite{Lifshitz}. Relevant experimental studies were being developed
for a long time \cite{old} until recently several high precision
measurements of the Casimir force were performed \cite{measure}.
Lifshitz theory and its application to various situations such 
as materials with finite conductivity and finite temperature effects 
has been an active area of research in the past few years \cite{klim}

The recent trend in miniaturization of mechanical devices naturally
brought up the issue that Casimir-Lifshitz interactions need to be
taken into consideration when small components are at close
proximity of each other \cite{nanomech}. For traditional design
strategies these forces, which could dominate all the others at
distances smaller than a few hundred nanometers, are to be
eliminated. On the other hand, one can also imagine using them for
novel design ideas that could potentially change the way we think
about designing mechanical systems at that scale \cite{machine}. For
these reasons, it is necessary to develop a better understanding of
Casimir-Lifshitz interactions when the objects involved do not have
ideal geometrical shapes.

This is far from a trivial task, but in recent years there have been
a number of significant developments to this end. These include
perturbative approaches for geometries that can be considered as
slightly deformed as compared to some ideal geometries
\cite{GK,EHGK,lambrecht}, semiclassical approaches \cite{semiclass}
and classical ray optics approximations \cite{Jaffe}, multiple
scattering and multipole expansions \cite{balian,klich,multipole1,multipole2,multipole3},
world-line method \cite{gies} and exact numerical diagonalization
method \cite{Emig-exact}, and the recently developed numerical
Green's function calculation method \cite{Johnson}. These different
and complementary approaches are useful in understanding the
subtleties involved in the dependence of the Casimir-Lifshitz energy
on geometrical features of the boundaries.

\begin{figure}
\includegraphics[width=.7\columnwidth]{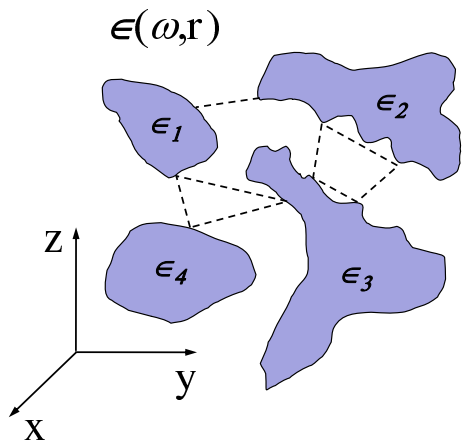}
\caption{Schematics of the dielectric function profile, and the
many-body contributions to the electromagnetic--fluctuation forces.}
\label{fig:schem1}
\end{figure}

An interesting result of the Lifshitz theory is that the interaction
at small separations are effectively determined by the value of
dielectric constants at relatively high (imaginary) frequencies
where it is not much different from unity \cite{Lifshitz}. This
suggests that a useful complementary strategy could be pursued based
on expansion in dielectric contrast. This approach, which has been
the subject of a few recent studies
\cite{barton,ramin,buhmann,rudi,milton}, is useful as it can treat the
effect of the geometry of boundaries exactly. Here we present a
systematic formulation of this approach for the calculation of
Casimir-Lifshitz energy for dielectric objects of arbitrary shape,
in the form of an expansion in powers of the dielectric contrast in
the medium (see Fig. \ref{fig:schem1}). We find that expansion in
powers of the difference between dielectric constant as compared to
the background takes on the form of an expansion in multi-body
contributions to the interaction. We provide explicit expressions
for each term in the expansion in the form of convolutions of a
tensorial kernel. We show that a resummation of the expansion can
help augment the convergence properties of the series, and make it
applicable to a wider range of dielectric properties.

The rest of the paper is organized as follows. Section
\ref{sec:formal} lays out the general formalism used for calculating
the Casimir-Lifshitz interaction, which is applicable in general to
metals and dielectrics. In Sec. \ref{sec:diel}, the dielectric
contrast perturbation theory is developed based on the formalism of
Sec. \ref{sec:formal} and an explicit expression is obtained for
each term in the expansion. In Sec. \ref{sec:CM}, the series
obtained in Sec. \ref{sec:diel} is resummed using a decomposition of
the kernel involved in the general expression for the energy. Section \ref{sec:CP}
gives the derivation of the Casimir-Polder energy using the formalism developed in
Sec. \ref{sec:CM}, as an example. The convergence properties of the series obtained as well as the nature of the divergent contributions in the theory are discussed in Sec.
\ref{sec:conv}. Section \ref{sec:parallel} is devoted to a specific
class of geometries where two nearly parallel semi-infinite
dielectric objects are placed in front of each other. Finally, Sec.
\ref{sec:disc} concludes the paper with some discussions.

\section{General Formalism} \label{sec:formal}

Let us assume that we have an assortment of dielectric objects in
space with arbitrary shapes and frequency dependent dielectric
properties, as sketched in Fig. \ref{fig:schem1}. This medium can be
described using a frequency- and space-dependent dielectric function
$\epsilon(\omega,{\bf r})$. We consider fluctuating electromagnetic
fields in this medium, which we choose to describe in the temporal
gauge, where the electrostatic potential vanishes identically,
namely $\phi=0$.

We start from the thermal (finite temperature) Green's function of
the electromagnetic field ${\cal D}_{ij}(\zeta_s;{\bf r},{\bf r}')$
in imaginary frequency, where $\zeta_s=2\pi s k_{\rm B} T/\hbar$ is
the Matsubara frequency with $k_{\rm B} T$ being the thermal energy
and $s$ a positive integer \cite{LanLif}. In the medium described
above, the thermal Green function is the solution to the following
equation
\begin{equation}
\left[\frac{\zeta_s^2}{c^2} \epsilon(i \zeta_s,{\bf
r})\delta_{ij}+\partial_i
\partial_j-\partial_l^2 \delta_{ij}\right]
{\cal D}_{jk}(\zeta_s;{\bf r},{\bf r}')=\delta_{ik}\delta^3({\bf
r}-{\bf r}').\label{eq:therm-green-def}
\end{equation}
By introducing the operator
\begin{equation}
{\cal K}_{ij}(\zeta_s;{\bf r},{\bf r}')=\left[\frac{\zeta_s^2}{c^2}
\epsilon(i \zeta_s,{\bf r})\delta_{ij}+\partial_i
\partial_j-\partial_l^2 \delta_{ij}\right] \delta^3({\bf
r}-{\bf r}'), \label{Kij-1}
\end{equation}
we can write Eq. (\ref{eq:therm-green-def}) as
\begin{equation}
\int d^3 {\bf r}_1 \;{\cal K}_{ij}(\zeta_s;{\bf r},{\bf r}_1) {\cal
D}_{jk}(\zeta_s;{\bf r}_1,{\bf r}')=\delta_{ik}\delta^3({\bf r}-{\bf
r}'),\label{eq:green-int-def}
\end{equation}
which means that as an operator we have
\begin{equation}
{\cal K}_{ij}(\zeta_s;{\bf r},{\bf r}')={\cal
D}_{ij}^{-1}(\zeta_s;{\bf r},{\bf r}').\label{eq:op-def}
\end{equation}
Now imagine a process in which we start from the empty space and
introduce the dielectric objects into the space in a perturbative
manner, similar to the charging process of a capacitor or
constructing a charge distribution by bringing infinitesimal charge
elements from infinity to assemble the distribution. Using standard
diagrammatic techniques \cite{Dzy}, we can show that the
introduction of the dielectric objects causes the Helmholtz free
energy of the system to change according to the following formula
\cite{LanLif}
\begin{widetext}
\begin{equation}
\delta {\cal F}=\delta {\cal F}_0+k_{\rm B} T
{\sum_{s=0}^{\infty}}^\prime \int d^3 {\bf r}_1 d^3 {\bf r}_2
\;{\cal D}_{ij}(\zeta_s;{\bf r}_1,{\bf r}_2) \delta {\cal
P}_{ji}(\zeta_s;{\bf r}_2,{\bf r}_1),\label{eq:helm-del-def}
\end{equation}
where
\begin{equation}
\delta {\cal P}_{ij}(\zeta_s;{\bf r},{\bf r}')=\frac{\zeta_s^2}{c^2}
\delta_{ij} \delta^3({\bf r}-{\bf r}') \delta \epsilon(i
\zeta_s,{\bf r}),\label{eq:polar-def}
\end{equation}
is the Polarization operator, and the primed summation means that
the $s=0$ term has an extra factor of $\frac{1}{2}$. In the
diagrammatics, the Polarization operator is formally related to the
Green function via the Dyson equation
\begin{equation}
{\cal D}_{ij}(\zeta_s;{\bf r},{\bf r}')={\cal
D}_{ij}^{(0)}(\zeta_s;{\bf r},{\bf r}')-\int d^3 {\bf r}_1 d^3 {\bf
r}_2 \; {\cal D}_{ik}^{(0)}(\zeta_s;{\bf r},{\bf r}_1) \delta {\cal
P}_{kl}(\zeta_s;{\bf r}_1,{\bf r}_2) {\cal D}_{lj}(\zeta_s;{\bf
r}_2,{\bf r}') ,\label{eq:dyson-def}
\end{equation}
\end{widetext}
where ${\cal D}_{ij}^{(0)}(\zeta_s;{\bf r},{\bf r}')$
denotes the Green function before the change in the dielectric
profile due to the introduction of new material. Equation
(\ref{eq:dyson-def}) can be solved to yield
\begin{equation}
{\cal D}_{ij}^{-1}(\zeta_s;{\bf r},{\bf r}')={{\cal
D}_{ij}^{(0)}}^{-1}(\zeta_s;{\bf r},{\bf r}')+\delta {\cal
P}_{ij}(\zeta_s;{\bf r},{\bf r}').\label{eq:dyson-inv}
\end{equation}
Equations (\ref{Kij-1}) and (\ref{eq:polar-def}) lead us to the following 
observation
\begin{equation}
\delta {\cal P}_{ij}(\zeta_s;{\bf r},{\bf r}')=\delta {\cal
K}_{ij}(\zeta_s;{\bf r},{\bf r}').\label{eq:dpol-dk}
\end{equation}
which helps us to write Eq. (\ref{eq:helm-del-def}) in the form of
\begin{equation}
\delta {\cal F}=\delta {\cal F}_0+k_{\rm B} T {\sum_{s=0}^{\infty}}
^\prime {\rm tr}  \left[{\cal K}^{-1} \delta {\cal K}
\right].\label{eq:helm-del-k}
\end{equation}
Equation (\ref{eq:helm-del-k}) can be formally integrated to yield
the following expression for the contribution to the Helmholtz free
energy due to Casimir-Lifshitz (CL) interactions:
\begin{equation}
{\cal F}_{\rm CL}=k_{\rm B} T {\sum_{s=0}^{\infty}}^\prime  {\rm tr}
\ln \left[{\cal K}_{ij}(\zeta_s;{\bf r},{\bf
r}')\right].\label{eq:helm-CL}
\end{equation}
At (effectively) low temperatures, we can convert the summation over
$s$ into an integration by changing $\Delta s=1$ into $d s=\hbar d
\zeta/(2 \pi k_{\rm B} T)$. In this case, we find the
Casimir-Lifshitz energy as
\begin{equation}
E_{\rm CL}=\hbar \int_0^\infty \frac{d \zeta}{2 \pi} \; {\rm tr} \ln
\left[{\cal K}_{ij}(\zeta;{\bf r},{\bf
r}')\right].\label{eq:E-CL-def}
\end{equation}
This could be a relatively simple starting point for the calculation
of the Casimir-Lifshitz energy in any system.

\section{Dielectric Contrast Perturbation Theory}   \label{sec:diel}

We can now construct a systematic perturbation theory scheme based
on the above definition of the Casimir-Lifshitz energy. The starting
point is to write the dielectric function profile as
\begin{equation}
\epsilon(i \zeta,{\bf r})=1+\delta \epsilon(i \zeta,{\bf
r}),\label{eq:eps-def}
\end{equation}
which can be used to decompose the kernel ${\cal K}_{ij}$ into a
part that corresponds to the empty space and a perturbation that
entails the dielectric inhomogeneity profile. In Fourier space, this
reads
\begin{equation}
{\cal K}_{ij}(\zeta;{\bf q},{\bf q}')={\cal K}_{0,ij}(\zeta,{\bf q})
(2 \pi)^3 \delta^3({\bf q}+{\bf q}')+\delta {\cal K}_{ij}(\zeta;{\bf
q},{\bf q}'),\label{eq:K=K0+delK}
\end{equation}
where
\begin{equation}
{\cal K}_{0,ij}(\zeta,{\bf q})=\frac{\zeta^2}{c^2} \delta_{ij}+q^2
\delta_{ij}-q_i q_j,\label{eq:K0-def}
\end{equation}
and
\begin{equation}
\delta {\cal K}_{ij}(\zeta;{\bf q},{\bf q}')=\frac{\zeta^2}{c^2}
\delta_{ij} \delta \tilde{\epsilon}(i \zeta,{\bf q}+{\bf
q}').\label{eq:delK-def}
\end{equation}
This decomposition can now be used to construct the perturbation
theory.

The expressions for the Casimir-Lifshitz energy [Eqs.
(\ref{eq:helm-CL}) and (\ref{eq:E-CL-def})] involve ${\rm tr} \ln
\left[{\cal K}\right]$, which can be written as a perturbative
series by using
\begin{eqnarray}
{\rm tr} \ln \left[{\cal K}\right]&=&{\rm tr} \ln \left[{\cal
K}_0\right]+{\rm tr} \ln [{\cal I}+{\cal K}_0^{-1} \delta {\cal
K}] \nonumber \\
&=&{\rm tr} \ln \left[{\cal K}_0\right]+\sum_{n=1}^{\infty}
\frac{(-1)^{n-1}}{n} \; {\rm tr}\left[\left({\cal K}_0^{-1} \delta
{\cal K}\right)^n\right],\nonumber \\
\label{eq:trlnK}
\end{eqnarray}
where ${\cal I}$ is the identity tensor and
\begin{equation}
{\cal K}_{0,ij}^{-1}(\zeta,{\bf q})=\frac{\frac{\zeta^2}{c^2}
\delta_{ij}+q_i q_j}{\frac{\zeta^2}{c^2}
\left[\frac{\zeta^2}{c^2}+q^2\right]}.
\end{equation}
Putting in the explicit forms for ${\cal K}_{0}^{-1}$ and $\delta
{\cal K}$, we can write the explicit form for the trace as
\begin{widetext}
\begin{equation}
{\rm tr}[({\cal K}_0^{-1} \delta {\cal K})^n]=\int \frac{d^3 {\bf
q}^{(1)}}{(2 \pi)^3} \cdots \frac{d^3 {\bf q}^{(n)}}{(2
\pi)^3}\frac{[\frac{\zeta^2}{c^2} \delta_{i_{1}
i_{2}}+q_{i_{1}}^{(1)} q_{i_{2}}^{(1)}] \cdots [\frac{\zeta^2}{c^2}
\delta_{i_{n} i_{1}}+q_{i_{n}}^{(n)}
q_{i_{1}}^{(n)}]}{[\frac{\zeta^2}{c^2} +q^{(1)2}] \cdots
[\frac{\zeta^2}{c^2} +q^{(n)2}]}\delta \tilde{\epsilon}(i
\zeta,-{\bf q}^{(1)}+{\bf q}^{(2)}) \cdots \; \delta
\tilde{\epsilon}(i \zeta,-{\bf q}^{(n)}+{\bf q}^{(1)}).
\label{trKKn-Fourier}
\end{equation}
The above expression contains the geometric information about the
arrangement of the dielectric objects through the Fourier transform
of the dielectric function profile. We can now rewrite Eq.
(\ref{trKKn-Fourier}) in real space, and find the following series
expression for the Casimir-Lifshitz energy of any heterogeneous
dielectric medium
\begin{equation}
E_{\rm CL}=\hbar \int_0^\infty \frac{d \zeta}{2 \pi}
\;\sum_{n=1}^{\infty} \frac{(-1)^{n-1}}{n} \;\int d^3 {\bf r}_1
\cdots d^3 {\bf r}_n \;{\cal G}^{\zeta}_{i_{1}i_{2}}({\bf r}_1-{\bf
r}_2) \cdots \; {\cal G}^{\zeta}_{i_{n}i_{1}}({\bf r}_n-{\bf
r}_1)\;\delta \epsilon(i \zeta,{\bf r}_1) \cdots \; \delta
\epsilon(i \zeta,{\bf r}_n), \label{E-gen}
\end{equation}
where the Green function defined as ${\cal G}^{\zeta}_{ij}({\bf
r})=\frac{\zeta^2}{c^2}{\cal K}_{0,ij}^{-1}(\zeta,{\bf r})$ reads
\begin{equation}
{\cal G}^{\zeta}_{ij}({\bf r})=\frac{\zeta^2}{c^2} \frac{{\rm
e}^{-\zeta r/c}}{4 \pi r} \left[\delta_{ij} \left(1+\frac{c}{\zeta
r}+\frac{c^2}{\zeta^2 r^2}\right)-\frac{r_i r_j}{r^2} \left(1+3
\frac{c}{\zeta r}+3 \frac{c^2}{\zeta^2
r^2}\right)\right]+\frac{1}{3} \delta_{ij} \delta^3({\bf r}),
\label{Gij-1}
\end{equation}
\end{widetext}
in real space. The tensorial kernel ${\cal G}^{\zeta}_{ij}$, has the
structure of the electric field of a radiating dipole in imaginary
frequency, including the delta function contribution that ensures
appropriate behavior of the expression in the near-field
\cite{Jackson}. This kernel has been introduced some time ago in
connection with van der Waals interactions \cite{Green}.

The main result of Eq. (\ref{E-gen}) [and a re-summed version of it
given in Eq. (\ref{E-gen-2}) below] has a number of interesting
characteristics. First, We find that an expansion in powers of
$\delta \epsilon$ automatically turns into a summation of integrated
contributions of $n$-body interactions. Moreover, all of the
$n$-body interaction terms have simple explicit expressions in terms
of a single fundamental kernel that mediates the two-body part of
the interaction.  Finally, we find a closed form expression for the
Lifshitz energy for any geometrical arrangement of dielectric bodies
in terms of quadratures, which is amenable to simple diagrammatic
rules and can hence be easily adapted for numerical computations at
any given order of the perturbation theory.

\section{Clausius-Mossotti Resummation of the Perturbation Theory} \label{sec:CM}

The perturbation theory developed above uses the spatial contrast in
the dielectric function as expansion parameter and this in general
may not be a suitably controlled expansion parameter, especially at
zero frequency where the dielectric contrast could be considerably
larger than unity (see Sec. \ref{sec:conv} for more discussion). The
formulation can be augmented by resumming the perturbation theory
such that it is organized as an expansion in powers of the
combination $\frac{\delta \epsilon}{1+\delta \epsilon/3}=\frac{3
(\epsilon-1)}{(2+\epsilon)}$ instead of $\delta \epsilon$. The new
expansion parameter, which reminds us of the Clausius-Mossotti
equation for molecular polarizability \cite{Jackson}, is
systematically smaller than the dielectric contrast itself, and is
finite even for real metals at zero frequency (where the dielectric
contrast diverges).

The key to this remedy lies in the Dirac delta function term in the
expression of the kernel in Eq. (\ref{Gij-1}). This suggests a
decomposition of the form
\begin{equation}
{\cal K}_{0}^{-1}=\frac{c^2}{3 \zeta^2} \left[3 {\cal A}+{\cal
I}\right],\label{eq:K0-1AI}
\end{equation}
to be used in the formulation, where the operator ${\cal A}$ is
defined as
\begin{equation}
{\cal A}_{ij}(\zeta,{\bf q})=\frac{2 \frac{\zeta^2}{c^2}
\delta_{ij}+3 q_i q_j-q^2 \delta_{ij} }{3
\left[\frac{\zeta^2}{c^2}+q^2\right]},\label{eq:Aq-def}
\end{equation}
in Fourier space, and as
\begin{eqnarray}
{\cal A}_{ij}(\zeta,{\bf r})&=&\frac{\zeta^2}{c^2} \frac{{\rm e}^{-\zeta r/c}}{4 \pi r}
\left[\delta_{ij} \left(1+\frac{c}{\zeta r}+\frac{c^2}{\zeta^2
r^2}\right) \right.\nonumber \\
&&\left.-\frac{r_i r_j}{r^2} \left(1+3 \frac{c}{\zeta r}+3
\frac{c^2}{\zeta^2 r^2}\right)\right], \label{eq:Ar-def}
\end{eqnarray}
in position space. Putting Eq. (\ref{eq:K0-1AI}) in Eq.
(\ref{eq:trlnK}), we find
\begin{eqnarray}
{\rm tr} \ln \left[{\cal K}\right]&=&{\rm tr} \ln \left[{\cal
K}_0\right]+{\rm tr} \ln
\left[{\cal I}+{\cal K}_0^{-1} \delta {\cal K}\right],\nonumber \\
&=&{\rm tr} \ln \left[{\cal K}_0\right]+{\rm tr} \ln
\left[\left({\cal I}+\frac{c^2}{3 \zeta^2} {\cal I} \delta {\cal K}
\right)
+\frac{c^2}{\zeta^2} {\cal A} \delta {\cal K}\right],\nonumber \\
&=&{\rm tr} \ln \left[{\cal K}_0\right]+{\rm tr} \ln \left[{\cal
I}+\frac{c^2}{3 \zeta^2} \delta {\cal K}\right]\nonumber \\
&&+{\rm tr} \ln \left[{\cal I}+\frac{c^2}{\zeta^2} {\cal A} \delta
{\cal K} \left({\cal I}+\frac{c^2}{3 \zeta^2} \delta {\cal K}
\right)^{-1}\right].\label{eq:trlnK2}
\end{eqnarray}
Defining a new operator
\begin{equation}
\delta {\cal B} \equiv \frac{c^2}{\zeta^2} \delta {\cal K}
\left({\cal I}+\frac{c^2}{3 \zeta^2}\delta {\cal K} \right)^{-1},
\end{equation}
which in position space has the following explicit form
\begin{equation}
\delta {\cal B}_{ij}(\zeta;{\bf r},{\bf r}')=\left[\frac{\delta
\epsilon(i \zeta,{\bf r})}{1+\frac{1}{3} \delta \epsilon(i
\zeta,{\bf r}) }\right] \delta_{ij} \delta^3({\bf r}-{\bf
r}'),\label{eq:dB-def}
\end{equation}
we can rewrite Eq. (\ref{eq:trlnK2}) as
\begin{eqnarray}
{\rm tr} \ln \left[{\cal K}\right]&=&{\rm tr} \ln \left[{\cal
K}_0\right]+{\rm tr} \ln \left[{\cal I}+\frac{c^2}{3 \zeta^2} \delta
{\cal K} \right]+{\rm tr} \ln \left[{\cal I}
+{\cal A} \delta {\cal B}\right],\nonumber \\
&=&{\rm tr} \ln \left[{\cal K}_0\right]+{\rm tr} \ln \left[{\cal
I}+\frac{c^2}{3 \zeta^2} \delta {\cal K} \right]\nonumber \\
&&+\sum_{n=1}^{\infty} \frac{(-1)^{n-1}}{n} \; {\rm
tr}\left[\left({\cal A} \delta {\cal
B}\right)^n\right].\label{eq:trlnK3}
\end{eqnarray}
This form can now be used to recast the perturbative expansion for
the Casimir-Lifshitz energy as a systematic expansion in powers of
the new perturbation parameter, which is always less than unity. We
thus find
\begin{eqnarray}
E_{\rm CL}&=&\hbar \int_0^\infty \frac{d \zeta}{2 \pi}
\;\sum_{n=1}^{\infty} \frac{(-1)^{n-1}}{n} \int
d^3 {\bf r}_1 \cdots d^3 {\bf r}_n \nonumber \\
&\times&{\cal A}_{i_{1}i_{2}}({\bf r}_1-{\bf r}_2) \cdots {\cal
A}_{i_{n}i_{1}}({\bf r}_n-{\bf r}_1)
\nonumber \\
&\times&\left[\frac{\delta \epsilon(i \zeta,{\bf
r}_1)}{1+\frac{1}{3} \delta \epsilon(i \zeta,{\bf r}_1) }\right]
\cdots  \left[\frac{\delta \epsilon(i \zeta,{\bf
r}_n)}{1+\frac{1}{3} \delta \epsilon(i \zeta,{\bf r}_n) }\right],
\label{E-gen-2}
\end{eqnarray}
in real space. In this result, we have neglected two remaining terms
in Eq. (\ref{eq:trlnK3}). The first one is the trivial term ${\rm
tr} \ln \left[{\cal K}_0\right]$ that corresponds to the self energy
of vacuum and could be easily eliminated as it does not depend on
any of the physical parameters involved. The second term in Eq.
(\ref{eq:trlnK3}) is a singular contribution, which has the explicit
form of
\begin{equation}
\hbar \int_0^\infty \frac{d \zeta}{2 \pi} \int d^3 {\bf r} \; \ln
\left[1+\frac{1}{3}\delta \epsilon(i \zeta,{\bf r})\right] \int
\frac{d^3 {\bf q}}{(2 \pi)^3}. \label{E-sing-1}
\end{equation}
This term does depend on the geometry of the dielectric objects, and
need to be subtracted off for any given geometry and dielectric
configuration. We will comment on the general issues related to such
divergent contributions as well as the convergence properties of the
series in the following Section.

\section{Example: Casimir-Polder Interaction}
\label{sec:CP}

To see how the formalism works, let us use it to calculate the
Casimir-Polder (CP) interaction between two objects at long
separations. Consider a sphere of volume $v_1$ and dielectric function
$\epsilon_1(i \zeta)$ located at the origin, and a sphere of volume
$v_2$ and dielectric function $\epsilon_2(i \zeta)$ located at a separation $R$
from the first one. To calculate the Casimir-Polder energy, we can use
Eq. (\ref{E-gen-2}) in conjunction with
\begin{eqnarray}
\frac{\delta \epsilon(i \zeta,{\bf
r})}{1+\frac{1}{3} \delta \epsilon(i \zeta,{\bf r})}&=&\frac{\delta \epsilon_1(i \zeta)}{1+\frac{1}{3} \delta \epsilon_1(i \zeta)} ~v_1 \delta^3({\bf r})\nonumber \\
&+&\frac{\delta \epsilon_2(i \zeta)}{1+\frac{1}{3} \delta \epsilon_2(i \zeta)} ~ v_2 \delta^3({\bf r}-{\bf R}).\label{epsilon-CP}
\end{eqnarray}
This is a singular form for the profile and is only valid in the limit
that the separation is much larger than the typical size of the objects.
Therefore, to be consistent with this singular limit we should only keep the leading order term in the size of the objects. A typical term in Eq. (\ref{E-gen-2}) with this dielectric profile, which does not contribute to the self energies of the two objects,
would (symbolically) look like
\begin{eqnarray}
&&\left(\frac{\delta \epsilon_1(i \zeta)}{1+\frac{1}{3} \delta \epsilon_1(i \zeta)}\right)^m ~v_1^m \left(\frac{\delta \epsilon_2(i \zeta)}{1+\frac{1}{3} \delta \epsilon_2(i \zeta)}\right)^{n-m} ~v_2^{n-m}\nonumber \\
&&\times~{\cal A}_{i_{1}i_{2}}({\bf 0}_1) \cdots {\cal
A}_{i_{m}i_{m+1}}(-{\bf R}) {\cal
A}_{i_{m+1}i_{m+2}}({\bf 0}_2)\cdots {\cal
A}_{i_{n}i_{1}}({\bf R}),\nonumber \\
\label{typ-CP}
\end{eqnarray}
which involves multiple powers of the quantity
\begin{equation}
f=v~{\cal A}_{ij}({\bf 0}).
\end{equation}
Using the proper definition of $f$ within our singular description
of the objects, we find
\begin{equation}
f=\lim_{{\bf r} \to {\bf 0}}~ \left[\left(\frac{4\pi}{3}r^3\right){\cal A}_{ij}({\bf r})\right]=0.\label{f}
\end{equation}
Therefore, to the leading order, we find
\begin{eqnarray}
E_{\rm CP}&=&-\hbar \int_0^\infty \frac{d \zeta}{2 \pi}~
\left[\frac{v_1\delta \epsilon_1(i \zeta)}{1+\frac{1}{3} \delta \epsilon_1(i \zeta)}\right] \left[\frac{v_2\delta \epsilon_2(i \zeta)}{1+\frac{1}{3} \delta \epsilon_2(i \zeta)}\right] \nonumber \\
&& \times ~{\cal A}_{ij}({\bf R}) {\cal A}_{ji}(-{\bf R}).\label{E-CP-1}
\end{eqnarray}
Using the following definition for dynamic polarizability
\begin{equation}
\alpha(i \zeta)=\frac{1}{4\pi}\left[\frac{v_1\delta \epsilon_1(i \zeta)}{1+\frac{1}{3} \delta \epsilon_1(i \zeta)}\right],\label{polar-def}
\end{equation}
and Eq. (\ref{eq:Ar-def}), we can rewrite Eq. (\ref{E-CP-1}) as
\begin{eqnarray}
E_{\rm CP}&=&-\frac{\hbar}{\pi R^6} \int_0^\infty d \zeta~ \alpha_1(i \zeta)
\alpha_2(i \zeta)~{\rm e}^{-2 \zeta R/c} \nonumber \\
&\times&\left[3+6 \frac{\zeta R}{c}+5 \frac{\zeta^2 R^2}{c^2}+2 \frac{\zeta^3 R^3}{c^3}+\frac{\zeta^4 R^4}{c^4}\right].\label{E-CP-2}
\end{eqnarray}
Ignoring the frequency dependence of the polarizabilities, we find
\begin{equation}
E_{\rm CP}=-\frac{23}{4 \pi}\frac{\hbar c}{R^7} \alpha_1
\alpha_2,\label{E-CP-3}
\end{equation}
which is the celebrated result for Casimir-Polder interaction \cite{CP}.
The above results could also be obtained using Eq. (\ref{E-gen}) in which case the
corresponding $f$-factors would create nonvanishing constant values that would need
to be added up to generate the Clausius-Mossotti form of the polarizabilities.

While the limiting form at longest separations can be obtained by this simple treatment, the correction terms (that are appreciable at closer separations) will systematically be produced
from to the $f$-factors that technically speaking correspond to higher multipole contributions to the Casimir-Polder energy, as well as many body contributions. To keep track of these corrections in a systematic way requires a multipole expansion type approach of the type developed in Refs. \cite{multipole1,multipole2,multipole3}. Finally, we note that the simple treatment given above corresponds to spheres where the symmetry of the object easily guarantees that Eq. (\ref{f}) holds true. For more complicated geometries, the question of orientation as well as the shape of the object complicates matters more \cite{multipole1,multipole3} and a more systematic approach is needed.

\section{Convergence and Regularization of the Perturbation Theory}
\label{sec:conv}

\subsection{When is the series expansion convergent?}

\begin{figure}
\includegraphics[width=.9\columnwidth]{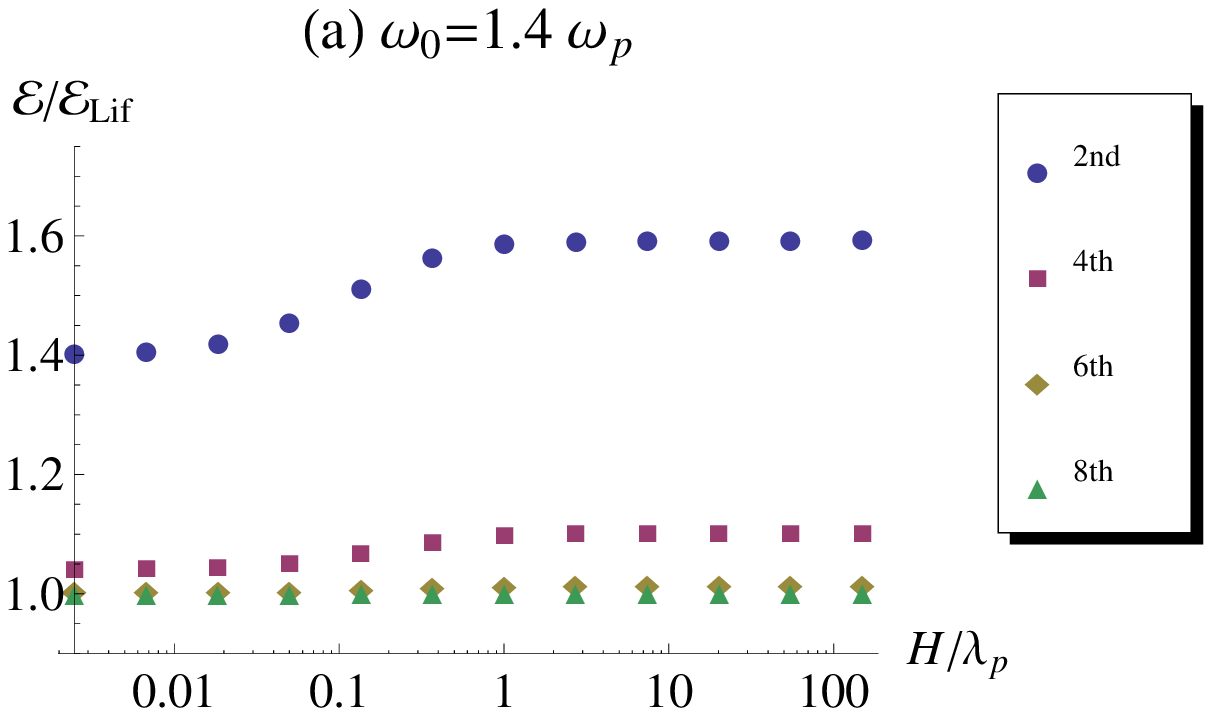}
\vskip0.7cm
\includegraphics[width=.9\columnwidth]{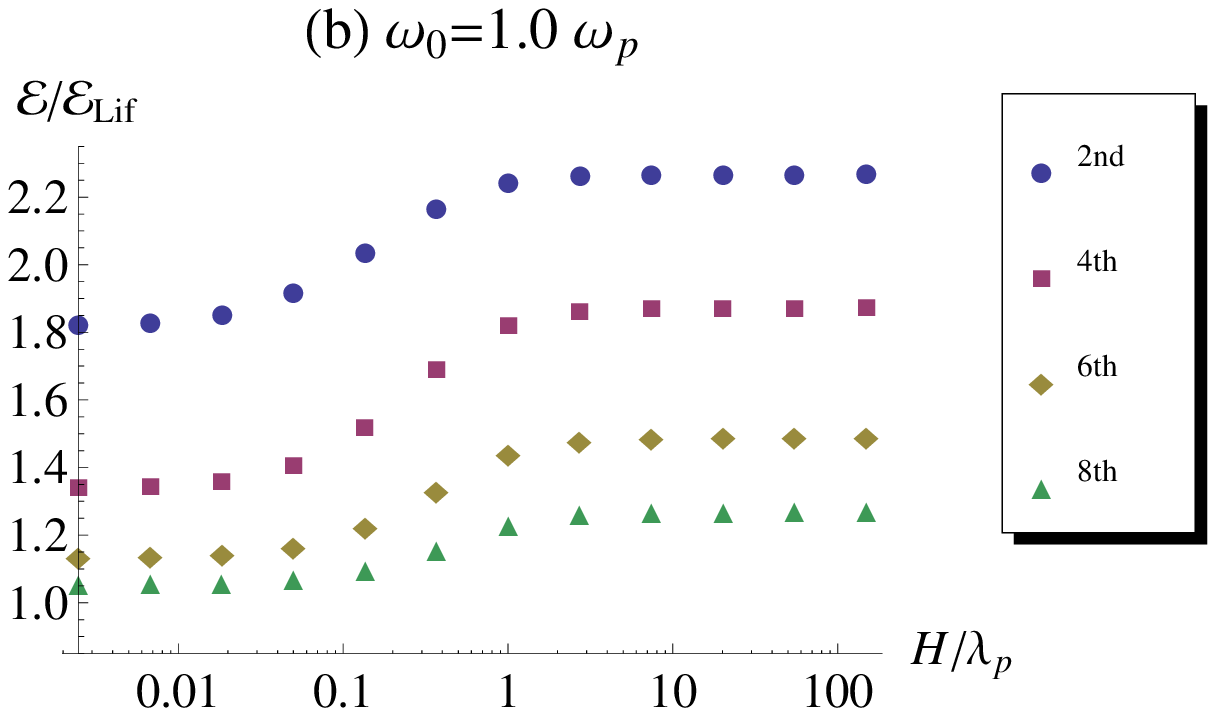}
\vskip0.7cm
\includegraphics[width=.9\columnwidth]{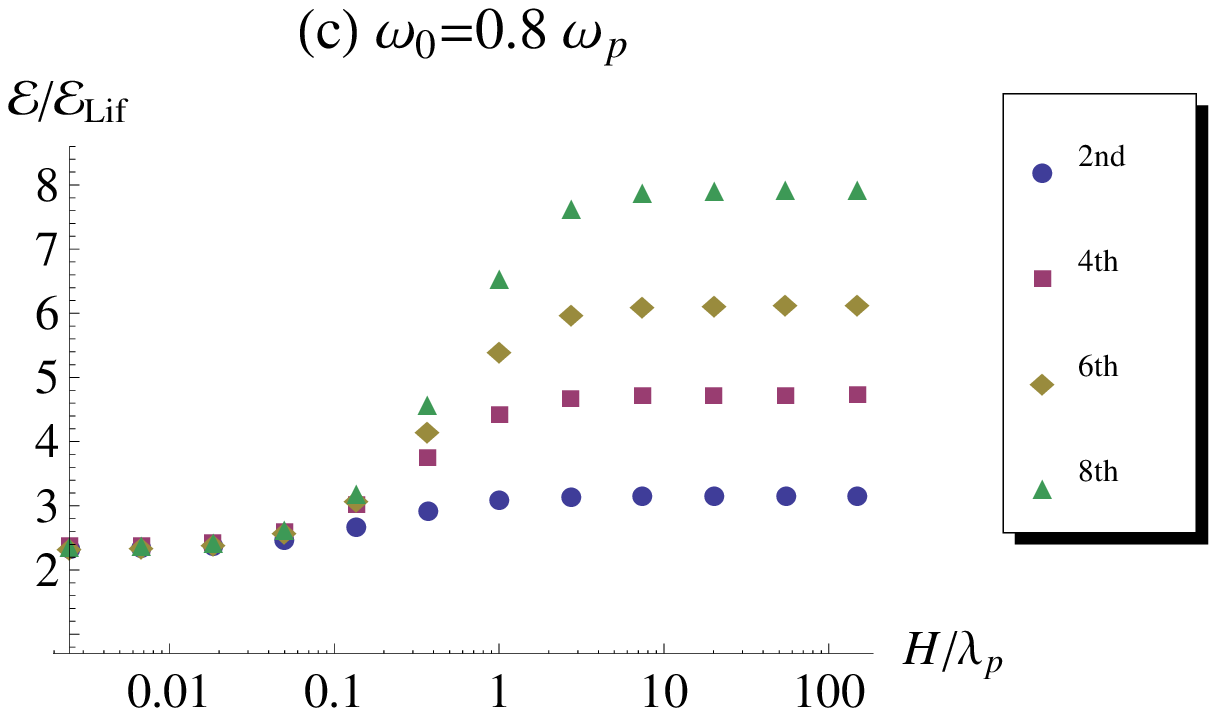}
\caption{Comparison of the expansion in Eq. (\ref{E-gen}) up to the
2nd, 4th, 6th, and 8th order, with the exact Lifshitz result for a
pair of parallel semi-infinite dielectrics, as a function of their
separation $H$ in units of the plasma wavelength $\lambda_{p}$. The
dielectric constant of both objects is chosen as Eq.
(\ref{eq:dielectric}) with (a) $\omega_0=1.4 \;\omega_{p}$, (b)
$\omega_0=1.0 \;\omega_{p}$, and (c) $\omega_0=0.8 \;\omega_{p}$.}
\label{fig:old}
\end{figure}

Because the formulation is constructed as a perturbation theory,
which can perhaps complement other perturbation theories for
Casimir-Lifshitz interaction between objects of arbitrary geometry,
we need to address the convergence properties of the expansion. To
examine the convergence property of Eq. (\ref{E-gen}), we can use a
specific example for which the exact Casimir-Lifshitz energy is
known, and compare it with the perturbative approximations at given
orders. We consider the case of two identical semi-infinite
dielectric objects with flat boundaries that are placed parallel to
each other at a separation $H$. The exact expression for the energy
per unit area for this problem is known to be \cite{Lifshitz}
\begin{eqnarray}
{\cal E}_{\rm Lif}&=&\frac{\hbar}{4 \pi^2 c^2} \int_0^{\infty} d
\zeta \; \zeta^2 \int_1^{\infty} d p \; p\nonumber \\
&&+\left\{\ln\left[1-\frac{(s-p)^2}{(s+p)^2} {\rm e}^{-2 p \zeta
H/c}\right]\right. \nonumber \\
&&+\left.\ln\left[1-\frac{(s-p \epsilon)^2}{(s+p \epsilon)^2} {\rm
e}^{-2 p \zeta H/c}\right]\right\},\label{eq:lif-flat}
\end{eqnarray}
where $s=\sqrt{\epsilon-1+p^2}$. We assume a simple form of
\begin{equation}
\epsilon(i
\zeta)=1+\frac{\omega_{p}^2}{\omega_0^2+\zeta^2},\label{eq:dielectric}
\end{equation}
for the dielectric constant (in imaginary frequency), where
$\omega_{p}$ represents the plasma frequency, from which the plasma
wavelength $\lambda_{p}=2 \pi c/\omega_{p}$ can be extracted. We
have checked that a dissipative term of the form $\gamma \zeta$ in
the denominator of Eq. (\ref{eq:dielectric}) would not alter are
results significantly for realistic values of $\gamma$, therefore we
have neglected this term for simplicity of our presentation.

\begin{figure}
\includegraphics[width=.9\columnwidth]{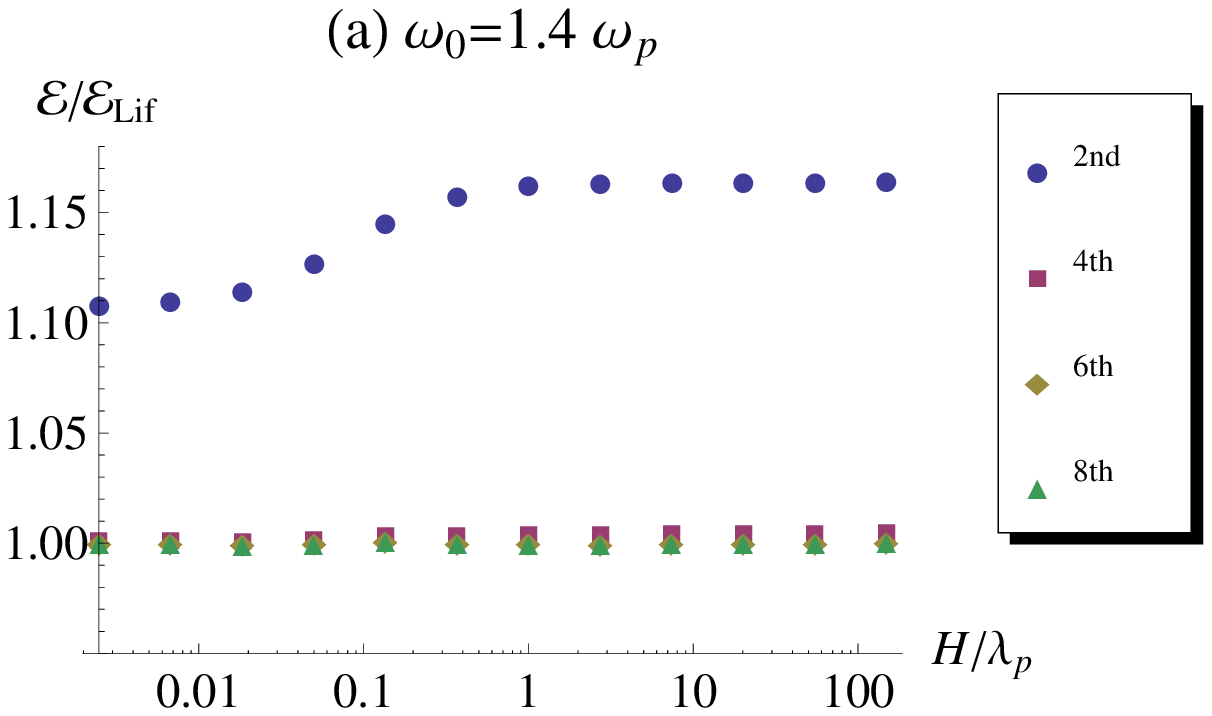}
\vskip0.7cm
\includegraphics[width=.9\columnwidth]{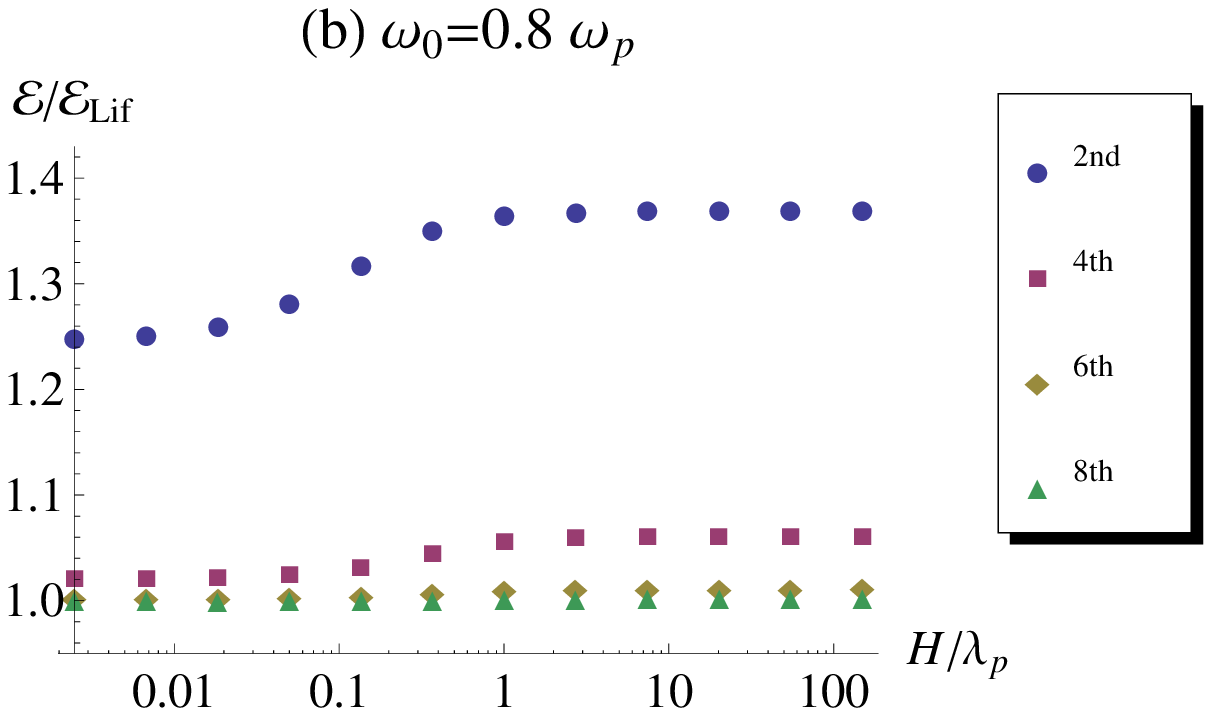}
\vskip0.7cm
\includegraphics[width=.9\columnwidth]{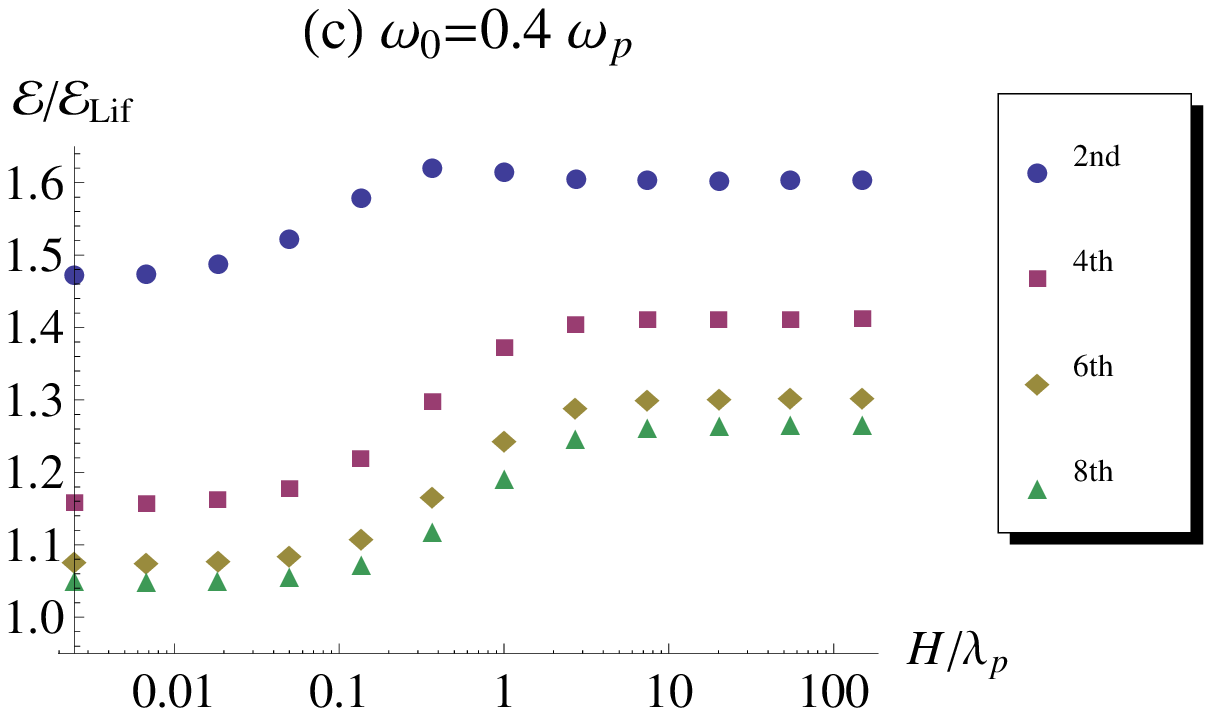}
\caption{Comparison of the expansion in Eq. (\ref{E-gen-2}) (the
so-called Clausius-Mossotti resummation) up to the 2nd, 4th, 6th,
and 8th order, with the exact Lifshitz result for a pair of parallel
semi-infinite dielectrics, as a function of their separation $H$ in
units of the plasma wavelength $\lambda_{p}$. The dielectric
constant of both objects is chosen as Eq. (\ref{eq:dielectric}) with
(a) $\omega_0=1.4 \;\omega_{p}$, (b) $\omega_0=0.8 \;\omega_{p}$,
and (c) $\omega_0=0.4 \;\omega_{p}$.} \label{fig:CM}
\end{figure}

In Fig. \ref{fig:old}, the ratio between the energy as calculated
from Eq. (\ref{E-gen}) and the exact Lifshitz result is shown as a
function of the separation, up to the second, fourth, sixth, and
eighth order in perturbation theory. The results clearly show a
crossover between two asymptotic regimes near $H=\lambda_{p}/(2
\pi)$ \cite{note1}. Figure \ref{fig:old}a corresponds to
$\omega_0=1.4 \;\omega_{p}$ and shows a rapid convergence. This can
be understood from the fact that even at zero frequency where the
dielectric contrast has its largest value, this example yields
$\delta \epsilon(0)=0.5$ that is smaller than unity. Figure
\ref{fig:old}b shows that when $\omega_0=1.0 \;\omega_{p}$ the
series is still convergent (despite $\delta \epsilon(0)=1$),
although not as rapidly as in the example of Fig. \ref{fig:old}a.
For $\omega_0=0.8 \;\omega_{p}$, the series is divergent, as Fig.
\ref{fig:old}c shows. Therefore, the series in Eq. (\ref{E-gen})
appears to be rapidly convergent when $\delta \epsilon(0)<1$, and
the convergence is considerably more efficient for $H <
\lambda_{p}/(2 \pi)$ [as compared to $H > \lambda_{p}/(2 \pi)$],
especially for $\delta \epsilon(0) \lesssim 1$.

The resummation of the perturbation theory discussed in Sec.
\ref{sec:CM} improves the convergence property of the series as it
effectively replaces $\delta \epsilon$ (that can have any value) by
$\frac{\delta \epsilon}{1+\delta \epsilon/3}=\frac{3
(\epsilon-1)}{(2+\epsilon)}$, which is bound to be between $0$ and
$3$ even for real metals at zero frequency. More explicitly, using
Eq. (\ref{eq:dielectric}) one can see that the expansion parameter
at zero frequency, which is the parameter that controls the
convergence of the series at large separations, changes from $\delta
\epsilon(0)=\omega_{p}^2/\omega_0^2$ to $3 \omega_{p}^2/(3
\omega_0^2+\omega_{p}^2)$. This suggests that at large separations,
the series [in Eq. (\ref{E-gen-2})] should now be convergent for
$\omega_0>0.8 \;\omega_{p}$ (so that the expansion parameter is
smaller than unity). Figure \ref{fig:CM} shows the convergence
property of a few examples using Eq. (\ref{E-gen-2}) instead of Eq.
(\ref{E-gen}). For $\omega_0=1.4 \;\omega_{p}$, a much more rapid
convergence is observed as shown in Fig. \ref{fig:CM}a, while the
previously divergent case of $\omega_0=0.8 \;\omega_{p}$ now shows a
good convergence, as seen in Fig. \ref{fig:CM}b and in agreement
with the argument above. As Fig. \ref{fig:CM}c shows, even for
$\omega_0=0.4 \;\omega_{p}$ one still observes a good convergence
despite the fact that the zero frequency expansion parameter is
equal to 2.

\begin{figure}
\includegraphics[width=.9\columnwidth]{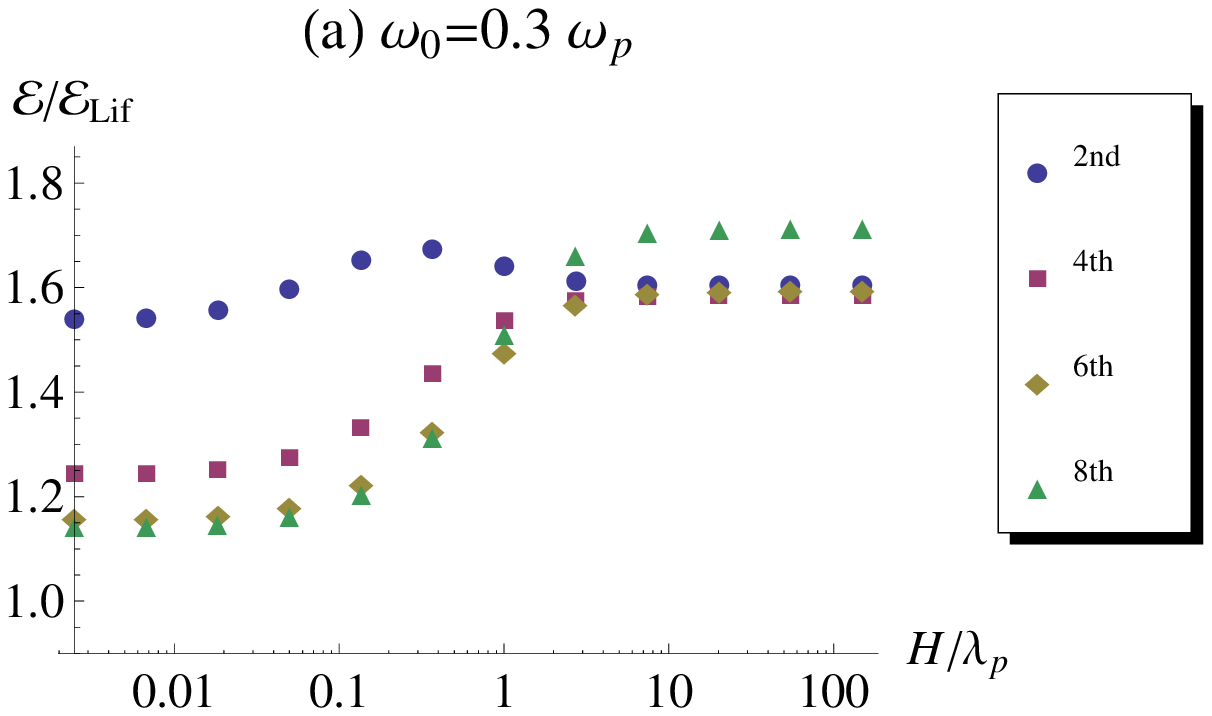}
\vskip0.7cm
\includegraphics[width=.9\columnwidth]{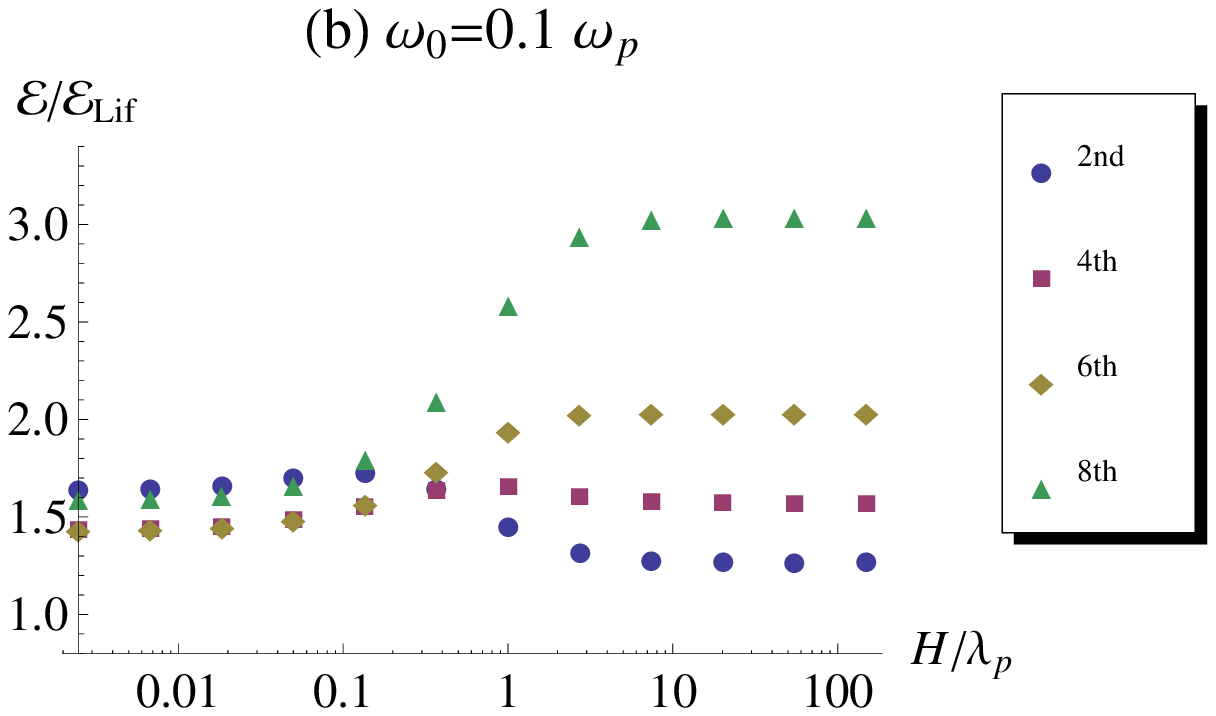}
\caption{Comparison of the expansion in Eq. (\ref{E-gen-2}) (the
so-called Clausius-Mossotti resummation) up to the 2nd, 4th, 6th,
and 8th order, with the exact Lifshitz result for a pair of parallel
semi-infinite dielectrics, as a function of their separation $H$ in
units of the plasma wavelength $\lambda_{p}$. The dielectric
constant of both objects is chosen as Eq. (\ref{eq:dielectric}) with
(a) $\omega_0=0.3 \;\omega_{p}$ (corresponding to silicon) and (b)
$\omega_0=0.1 \;\omega_{p}$.} \label{fig:div}
\end{figure}

Figure \ref{fig:div}a shows the result of the expansion for
$\omega_0=0.3 \;\omega_{p}$. Interestingly, it appears that in this
case the series is (at least up the eighth order in perturbation)
convergent for $H < \lambda_{p}/(2 \pi)$, while it clearly diverges
for $H > \lambda_{p}/(2 \pi)$. One can roughly say that the
appropriate parameter that controls the convergence at small
separations is the expansion parameter at $\zeta=\omega_0$, which is
$3 \omega_{p}^2/(6 \omega_0^2+\omega_{p}^2)$. For $\omega_0=0.3
\;\omega_{p}$, this parameter is $\simeq 2$, which is consistent
with the value of the zero frequency expansion parameter in the
borderline convergence case at large separations obtained for
$\omega_0=0.4 \;\omega_{p}$. Finally, for $\omega_0=0.1
\;\omega_{p}$, the series is divergent, as can be seen in Fig.
\ref{fig:div}b. It is interesting that the criterion for convergence
seems to be that the relevant expansion parameter---i.e. the
parameter at the relevant frequency---needs to be smaller than 2.

\subsection{Divergencies and Regularization}

A general issue with the calculation of the Casimir-Lifshitz energy
is the appearance of divergent contributions, like in any quantum
field theory calculation. In the case of a system with arbitrary
geometry, it is not clear how these divergencies depend on the
details of the geometry, so that they could be identified and dealt
with in a systematic way. The present formulation also suffers from
the presence of divergent terms, but similar to any perturbative
field theory with a manifest recipe for constructing each term,
these divergent contributions could be systematically identified and
regularized order by order in perturbation.

As pointed out by Barton \cite{barton}, it is important to examine
the nature of the divergent contributions, and for example determine
whether they are controlled by the minimum possible distance between
atoms and molecules (i.e. cutoff on wavevector), transparency of the
materials at high frequency (i.e. plasma frequency as cutoff) or
ultraviolet frequency cutoff in vacuum. This point has not always
been carefully dealt with in the literature of the Casimir-Lifshitz
interactions, partly because the largely used perfect conductor
limit already blurs this distinction at the outset when it assumes
the plasma frequency is infinite.

In our formulation, the divergencies originating from lack of
molecular excluded volume in the theory will be regularized using a
high wavevector cutoff in the ${\bf q}$ integrals in Eq.
(\ref{trKKn-Fourier}) or a short distance cutoff in the position
integrals in Eqs. (\ref{E-gen}) and (\ref{E-gen-2}). To see how this
works let us go to Eq. (\ref{trKKn-Fourier}) and look at the ${\bf
q}$ integrations. We have
\begin{eqnarray}
{\rm tr}[({\cal K}_0^{-1} \delta {\cal K})^n]=\int \frac{d^3 {\bf
q}^{(1)}}{(2 \pi)^3} &\cdots& \frac{d^3 {\bf q}^{(n)}}{(2
\pi)^3}~M({\bf q}^{(1)},\cdots,{\bf q}^{(n)})\nonumber \\
\times~ \delta \tilde{\epsilon}(i \zeta,-{\bf q}^{(1)}+{\bf
q}^{(2)}) &\cdots& \delta \tilde{\epsilon}(i \zeta,-{\bf
q}^{(n)}+{\bf q}^{(1)}). \label{trKKn-2}
\end{eqnarray}
where
\begin{eqnarray}
M({\bf q}^{(1)},&\cdots&,{\bf q}^{(n)})=\left[3 \zeta^{2 n}+\zeta^{2
n-2} \left(q^{(1)2}+\cdots+q^{(n)2}\right)\right.\nonumber \\
&&\left.+\cdots+\left({\bf q}^{(1)} \cdot {\bf q}^{(2)}\right)
\cdots \left({\bf q}^{(n)} \cdot {\bf q}^{(1)}\right) \right]/
\nonumber \\
&&\left[\left(\frac{\zeta^2}{c^2} +q^{(1)2}\right) \cdots
\left(\frac{\zeta^2}{c^2} +q^{(n)2}\right)\right].\label{eq:Mdef}
\end{eqnarray}
Since the (Fourier-space) dielectric contrast profiles only depend
on the differences between the wavevectors, we can change the
integration variables so that one independent wavevector can be
integrated out. Using ${\bf q}_i={\bf q}^{(i)}-{\bf Q}$ where ${\bf
Q}=\frac{1}{n}({\bf q}^{(1)}+\cdots+{\bf q}^{(n)})$, we find
\begin{eqnarray}
{\rm tr}[({\cal K}_0^{-1} \delta {\cal K})^n]=n \int \frac{d^3 {\bf
q}_{1}}{(2 \pi)^3} &\cdots& \frac{d^3 {\bf q}_{n-1}}{(2 \pi)^3}
\frac{d^3 {\bf Q}}{(2
\pi)^3} \nonumber \\
\times ~ M({\bf q}_1+{\bf Q},&\cdots&,{\bf q}_{n}+{\bf Q})\nonumber \\
\times~ \delta \tilde{\epsilon}(i \zeta,-{\bf q}_{1}+{\bf q}_{2})
&\cdots& \delta \tilde{\epsilon}(i \zeta,-{\bf q}_{n}+{\bf q}_{1}),
\label{trKKn-3}
\end{eqnarray}
where ${\bf q}_{n}=-{\bf q}_{1}-\cdots-{\bf q}_{n-1}$. The integral
\begin{equation}
\int \frac{d^3 {\bf Q}}{(2 \pi)^3}~M({\bf q}_1+{\bf Q},\cdots,{\bf
q}_{n}+{\bf Q}),\nonumber
\end{equation}
can now be performed independently, and looking at the definition of
the kernel $M$ in Eq. (\ref{eq:Mdef}) we can see that it diverges at
high values of ${\bf Q}$ where $M \to 1$. This divergent
contribution can be extracted by writing $M=M-1+1$ and separating
the $1$ factor, which leads to a regularized contribution
proportional to $\int \frac{d^3 {\bf Q}}{(2 \pi)^3}=\frac{1}{a^3}$.
The remaining ${\bf q}$ integrations for this divergent contribution
can be carried out, and the series can be summed up to yield an
overall singular contribution of the form
\begin{equation}
E_{\rm sing.}=\frac{\hbar}{a^3} \int_0^\infty \frac{d \zeta}{2 \pi}
\int d^3 {\bf r} \; \ln \left[\epsilon(i \zeta,{\bf r})\right],
\label{E-sing-2}
\end{equation}
which corresponds to volume terms and can be systematically isolated
for any geometry and dielectric configuration. This result entails a
similar volume contribution calculated in Ref. \cite{barton} at the
second order in $\delta \epsilon$. Other divergent contributions in
the present formulation can be dealt with in a similar manner to the
above example.

Finally, we note that using a very similar calculation
to the one presented in Sec. \ref{sec:CP} for the Casimir-Polder interaction,
one can show that for an arbitrary assortment as depicted in Fig. \ref{fig:schem1},
the Casimir-Lifshitz energy calculated from the present formalism 
in the limit where the objects are far from each other only contains divergent 
contributions in the self energies of these objects and all of the terms that depend
on the distances---including the many-body terms---are finite.

\section{Parallel Semi-Infinite Objects}    \label{sec:parallel}

\begin{figure}
\includegraphics[width=.6\columnwidth]{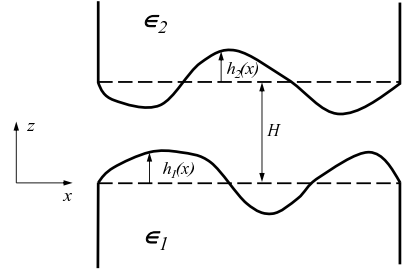}
\caption{Schematics of two semi-infinite dielectric bodies with
boundaries of non-ideal shape.} \label{fig:schem2}
\end{figure}

Due to its importance, we now focus our attention on the specific
arrangement shown in Fig. \ref{fig:schem2}, where two nearly
parallel semi-infinite dielectric bodies with irregularly shaped
boundaries are placed next to each other at a mean separation $H$.
We can write down the dielectric function profile in space as
\begin{equation}
\epsilon(i \zeta,{\bf r})=\left\{\begin{array}{ll}
\epsilon_2(i \zeta), & \; H+h_2({\bf x}) \leq z < +\infty,  \\ \\
1, & \; h_1({\bf x}) < z < H+h_2({\bf x}),  \\ \\
\epsilon_1(i \zeta), & \; -\infty < z \leq h_1({\bf x}),
\end{array} \right. \label{epsilon-profile}
\end{equation}
which in Fourier space reads
\begin{eqnarray}
\delta \tilde{\epsilon}(i \zeta,{\bf q})&=&\frac{i}{q_z} \int d^2
{\bf
x} \; {\rm e}^{i {\bf q}_{\perp} \cdot {\bf x}} \nonumber \\
&\times& ~\left[\delta \epsilon_2 \; {\rm e}^{i q_z \left[H+h_2({\bf
x})\right]}-\delta \epsilon_1 \; {\rm e}^{i q_z h_1({\bf
x})}\right]. \label{delta-epsilon-q}
\end{eqnarray}
When the parameters are such that the leading contribution in Eq.
(\ref{E-gen-2}) comes from the second order term (see Sec.
\ref{sec:conv}), we can simplify the expression for the
Casimir-Lifshitz interaction and write it in a closed form that can
be readily used in studies of various geometrical effects. A similar
strategy has been discussed in Ref. \cite{barton}, where a large
variety of other geometries has been considered. Putting in the
dielectric function profile of Eq. (\ref{epsilon-profile}), we find
\begin{eqnarray}
E_2&=&-\frac{9 \hbar}{128 \pi^3 c^4}\int_0^\infty d \zeta \; \zeta^4
\left[\frac{\epsilon_1(i \zeta)-1}{\epsilon_1(i
\zeta)+2}\right]\left[\frac{\epsilon_2(i \zeta)-1}{\epsilon_2(i
\zeta)+2}\right] \nonumber
\\
&\times& \int d^2 {\bf x} d^2 {\bf x}'\;{\cal
W}\left(\frac{\zeta}{c}[{\bf x}-{\bf
x}'],\frac{\zeta}{c}[H+h_2({\bf x})-h_1({\bf
x}')]\right),\nonumber \\
\label{E-2-1}
\end{eqnarray}
where
\begin{eqnarray}
{\cal W}({\bf y},h)=8 \;\Gamma\left(0,2
\sqrt{y^2+h^2}\right)+\int_{1}^{\infty} \frac{d s}{s^{3/2}}\;
{\rm e}^{-2 \sqrt{y^2+h^2 s}} \nonumber \\
\times\left[\frac{3}{(y^2+h^2 s)^2}+\frac{6}{(y^2+h^2
s)^{3/2}} +\frac{4 \;(1-h^2 s)}{(y^2+h^2 s)}\right],\nonumber \\
\label{Mxx'}
\end{eqnarray}
and $\Gamma(a,z)=\int_z^\infty d t \;t^{a-1}\;{\rm e}^{-t}$ is the
incomplete gamma function. Note that at this order, the Lifshitz
energy is pairwise additive, and that the effect of geometry has
been taken into account exactly for any arbitrary profile. In the
above result, we have kept the frequency dependence of the
dielectric functions as well as the geometry of the boundaries
arbitrary for generality of the presentation. The expression in Eq.
(\ref{E-2-1}) can be considerably simplified for $h_1({\bf x})=0$:
\begin{equation}
\left.E_2\right|_{h_1({\bf x})=0}=\int d^2 {\bf x}\; {\cal
E}_2\left(H+h_2({\bf x})\right),\label{E-2-h1=0}
\end{equation}
in terms of the original Lifshitz result for the energy per unit
area of flat boundaries (within the same scheme of Clausius-Mossotti
approximation) \cite{Lifshitz}
\begin{eqnarray}
{\cal E}_2(H)&=&-\frac{9 \hbar}{64 \pi^2 c^2}\int_0^\infty d \zeta
\; \zeta^2 \; {\cal L}\left(\frac{\zeta H}{c}\right)\nonumber
\\
&\times& \left[\frac{\epsilon_1(i \zeta)-1}{\epsilon_1(i
\zeta)+2}\right]\left[\frac{\epsilon_2(i \zeta)-1}{\epsilon_2(i
\zeta)+2}\right],\label{E-2-h1=0-2}
\end{eqnarray}
with
\begin{equation}
{\cal L}(u)=4 (u^2-1) E_1(2 u)+\frac{{\rm e}^{-2 u}}{u^2} (1+2
u+u^2-2 u^3),
\end{equation}
and the exponential integral function defined as
$E_n(z)=\int_1^\infty d t \;{\rm e}^{-z t}/t^n$. This simplification
is a general feature that is present for any pairwise additive
interactions, as has been shown in Ref. \cite{EHGK}.

\section{Discussion}    \label{sec:disc}

When dealing with complicated quantum field theories, it is always
helpful to try and formulate complementary perturbative schemes so
that the specific point of interest, which is usually out of reach,
is approached from different directions. This could potentially
provide complementary information that could be compiled to yield an
improved insight into the properties of the theory. A good example
of this synergy is the perturbation theory and $1/n$ expansion of
the $O(n)$ model of the $\phi^4$ quantum field theory \cite{zinn}. A
similar strategy has been the underlying motivation for the work
presented here: expansion in dielectric contrast allows us to
formulate a perturbative scheme for calculating the Casimir-Lifshitz
interaction between object with arbitrary geometry keeping the
effect of the geometry {\em exact}. The motivation behind this
alternative approach comes from the fact that at distances smaller
than the plasma wavelength the effective dielectric contrast that
determines the Casimir-Lifshitz interaction corresponds to the high
frequency limit of Eq. (\ref{eq:dielectric}), which is effectively
smaller than the equivalent value for the large distance regime.
This view is confirmed by the plots in Figs. \ref{fig:old},
\ref{fig:CM}, and \ref{fig:div}, which show that the series
expansion in dielectric contrast converges more rapidly at short
distances. Interestingly, Fig. \ref{fig:div}a suggests that it is
possible to have a convergence at distances smaller than the plasma
wavelength---typically a few hundred nanometers---while the same
series diverges at larger distances. This marginal case is observed
for $\omega_0=0.3 ~\omega_p$, which coincidentally corresponds to
silicon \cite{silicon} that is a common choice for fabrication of
small mechanical components with fine geometrical features.

The perturbation theory seems to have good convergence property for
materials that have a dielectric contrast of $\sim 2$, or a
dielectric constant of $\sim 3$, at the relevant frequency that is
zero for large distance asymptotics and $\omega_0$ for short
distance regime. While this is somewhat restrictive, as for example
it does not include real metals, it is not too far from being able
to include some commonly used dielectric materials, such as silicon
for example. Note that this complementary approach keeps the effect
of the geometrical features exact, which is a significant
improvement compared with complementary theories that apply to
objects with small deformations. Finally, we note that it might be
possible to use Borel summation method \cite{zinn} (or other
equivalent techniques) to improve the convergence of the series
expansion in the strong coupling limit of dielectric contrast.

In conclusion, we have developed a perturbative scheme for
calculating the Casimir-Lifshitz interaction between objects of
arbitrary geometry. Explicit expressions are provided for each term
in the perturbation theory, as multiple integrals over the bodies of
the interacting objects. This method could in principle be used to
calculate the interactions at any order using standard numerical
integration techniques.

\acknowledgments

This work was supported by EPSRC under Grant EP/E024076/1.


\begin{thebibliography}{99}
\bibitem{note1}
Note that the last few data points in Figs. \ref{fig:old},
\ref{fig:CM}, and \ref{fig:div} corresponding to large values of
$H/\lambda_{p}$ would lie in the thermal regime where the sum over
the frequency should be discretized [see Eq. (\ref{eq:helm-CL})].
These points are presented to help demonstrate the crossover
behavior in a convincing way.
\end{thebibliography}
\end{document}